\documentclass[twocolumn,showpacs,floatfix,preprintnumbers,amsmath,amssymb]{revtex4}
\usepackage[dvips]{graphicx} 
\usepackage{txfonts}
\usepackage{dcolumn}
\usepackage{bm}
\usepackage{epstopdf}
\usepackage[utf8]{inputenc} 
\begin{document} 

\preprint{APS}
\title{Bounce inflation with a conserved frame of rest}
%
\author{Gustaf Rydbeck}
\affiliation
{Onsala Space Observatory\\
Onsala 43497\\
Sweden
}%

%
%
\date{\today}
%
\begin{abstract}
 Some form of approximately exponential inflation is generally assumed to be the origin of our present universe. The inflation is thought to be driven by a scalar field potential where the field first slowly slides along the potential and then comes to a steep slope where the field rapidly falls and then oscillates around zero transforming into particles. The slowly sliding scalar field inflation leads to an exponentially expanding de Sitter space. A scalar field as well as the deSitter space are both Lorentz invariant. Thus no global frame of rest can be established in this scenario, while particle creation requires a preferred frame of rest. Observations of the cosmic microwave background show, when the redshift is corrected for our local velocity, a very even temperature and redshift distribution requiring a global preferred frame of rest. We suggest here that a density dependent equilibrium relation between matter/radiation and a scalar energy density could maintain a preferred frame of rest throughout the bounce and inflation and thereby solve the problem.
 \end{abstract}

\maketitle

\section{Introduction}
Some form of inflation is generally assumed to be the origin of the universe. A number of different inflation scenarios have been suggested, see e.g. \cite {porcelli} .The inflation is thought to be driven by the energy density of a scalar field. The scalar field can have a constant spatial value as well as wave-excitations representing particles. The only observed fundamental scalar field is the Higgs field  with the associated Higgs particle.The Higgs field could be inflation driving, see Javier Rubio(\cite {rubio} 2019), but is by no means the only possible inflaton field. Generally inflation scenarious leave three questions unanswered,
\begin{enumerate}
\item The physics and origin of the pre-inflation universe
\item The nature and physics of the inflation driving scalar field and it's potential.
\item Establishment of a  global frame of rest at the end of inflation.
\end{enumerate}
For the universe to be globally isothermal, the standard inflation scenario requires that the pre-inflation universe must have been thermalized for the scalar field to be globally isotropic and homogenous. For general initial conditions see \cite{dalia}. These conditions are however not sufficient  to establish a preferred global restframe at the end of inflation. The standard "cold" inflation would have dispersed all particles that existed before inflation. Since both the scalar field and the de Sitter space are Lorentz invariant there is no trace of a preferred reference frame. To keep a preferred restframe the inflation must necessarily be 'warm'. In contrast to most theories the "warm inflation" suggested by Berera 1995\cite{berera},\cite{berera2} and Graham and Moss 2009 \cite{graham}  is associated with a thermal heat bath which preserves a preferred global restframe. This still  leaves the question of the initial establishment of the preferred global restframe and the nature of the pre-inflation universe unanswered. \\ 
We suggested \cite{rydbeck} a different possibility that by introducing an upper density limit $\rho_m$, a collapsing space or stellar object forming a black hole would bounce into a new inflating space separated from the original space. We now suggest that by smoothing the approach to the upper density limit, a preferred thermal reference frame can be maintained throughout the bounce. This suggestion of course introduces an hypothetical  upper density limit, but a proposed scalarfield energy function, required by standard inflation,  is no less hypothetical, and we dont need to invent a pre inflation thermalized universe  with a magical creation. In addition our suggested upper density limit leads to a globally uniform temperature only perturbed by standard quantum fluctuations.
\section{the equation of state}
We will investigate an equation for the density dependance of scale as follows
\begin {equation}
\rho=\rho_0 \cdot \left(\frac{a_0}{a}\right)^{m(\rho)}
\label{scale}
\end{equation}
where $\rho$ is energy density and $a$ a scale factor.
$m(\rho)$, the scale index,  will have have different values for different types of the known energy densities
\begin{eqnarray} 
m(\rho_{s})=0  \\
m(\rho_{cu})=2 \nonumber \\
m(\rho_{nr})=3 \nonumber\\
m(\rho_r)=4     \nonumber
\end{eqnarray}
Where $s$ stand for scalar, $cu$ for curvature, $nr$ for nonrelativistic and $r$ for relativistc energy density.
  
 The differential form of  equation \ref{scale} is
\begin{equation}
\frac {d\rho} {\rho\; m(\rho)
\left(1-  \left( \frac{v_m}{c} \right) \right)}=-\frac{da}{a}
\label{model2}  
\end{equation}
where we have included a hysteresis effect, $1-  \frac{v_m}{c}$ where $v_m$ is the velocity at the eventhorizon distance $a_m$ of the limit density $\rho_m$ and c the speed of light (cf \cite{sayantan}). The hysteresis effect is an important ingredient in the equation since it gives a friction effect without which the contraction and the expansion would be symmetric, there would be no time evolution and no increase in entropy.
$\frac {v_m}{c} $ can, using the friedmann equation, be substituted by $ \mp \left( \frac {\rho_t}{\rho_m} \right)^{1/2} $  (- and + during contraction  and expansion  respectively ) where $\rho_t$ is the total density including the curvature density. We then have 
\begin{equation}
\frac {d\rho} {\rho \; m(\rho)
\left(1\pm  \left( \frac {\rho_t}{\rho_m} \right)^{1/2}\right)} =-\frac{da}{a}
\label{state}
\end{equation}
We will generally use units with the speed of light c=1. The pressure $p$ can be related to the density 
\begin{equation}
p=-\frac{dE}{dV}=-\frac{a}{3} \frac{d\rho}{da}-\rho= \rho \left(\frac{m(\rho)}{3} \left(1\pm \left( \frac{\rho_t}{\rho_m}\right)^{1/2}\right)-1\right)
\label{pre}
\end{equation}
We have set $\rho_m$ to $10^{-4} \cdot \rho_P$, where $\rho_P$ is the Planck density, which is a total shot in the dark but provides model results within observational constraints.\\
One might consider $\rho$ as a sum of two densities $\rho_s+\rho_r=\rho$, one with negative pressure $p_s=-\rho_s$ and one with radiation pressure  $p_r=1/3\rho_r$. The relation between $\rho$ and $\rho_r$ would then be
\begin{equation}
\rho_r=\rho \left(\frac{m(\rho)}{4} \left( 1 \pm  \left( \frac {\rho_t}{\rho_m} \right )^{1/2} \right)\right)
\end{equation}
$m(\rho)$ will have limits $0< m(\rho)\le 4$, ensuring that $\rho_r>0$ so that a preferred reference frame is always defined.
 \section{the Friedmann equation}
 In a limited region in the center of a collapsing star the density can be assumed to be constant. The evolution of this region can then be traced by the Friedmann equation.
 \begin{equation}
 \dot a^2={\cal{G}}\rho a^2 - k 
 \label{friedmann}
 \end{equation}
 in finite element form
 \begin{equation}
  \delta a=({\cal{G}}\rho a^2 - k)^{1/2} \delta t
  \label{fried}
 \end{equation}
 where $k=c^2$ is the curvature constant and ${\cal G}=\frac{8\pi}{3}G$ where $G$ is the gravitational constant. $\rho$ is the sum of relevant densities. In the present case it is 
 the sum of scalar or vacuum energy density and the relativistic  energy density (the sum of radiation and relativistic particles) . In our computer code for \ref{fried}, we can then as we step forward in time
 use equation \ref{state} to change the density accordingly. 
\section{quantum density fluctuations during inflation}
Relating energy density fluctuations to energy fluctuations
\begin{equation}
\Delta\rho=\frac{\Delta m}{\frac{4\pi}{3}   a^3}=\frac{\Delta E}{c^2   \frac{4\pi}{3}  a^3 }         
\end{equation}
Applying the uncertainty relation we have 
\begin{equation}
\Delta E \cdot \Delta T \ge \hbar/2
\end{equation}
$\Delta T$ is the time it takes for the fluctuation to expand to the event horizon ; $\Delta T_i=H_i^{-1}$, where $H_i$ is the Hubble constant at time $t_i$. The event horizon $a_H=\frac{c}{H } \Rightarrow$
\begin{equation}
\delta\rho_i=\frac{\hbar \, H_i}{2\, c^2\frac{4\pi}{3}\,a_{H_i}^3}=\frac{3\,\hbar \, H_i^4}{8\pi \, c^5 } \label{fluct2}
\end{equation}
Where $\delta \rho=\langle |\Delta \rho | \rangle$.
Assuming the curvature density is small compared to the  the relevant densities, Einstein-Friedman equations relate H to the density, 
\begin{equation}
 H_i=\left(\frac{8\pi}{3}\, G\rho_i \right) ^{1/2} 
\end{equation}
 Inserting in \ref{fluct2} we get
 \begin{equation}
 \delta \rho_i=\frac{\hbar 8\pi G^2 {\rho_i}^2}{3c^5}
 \end{equation}
 It follows that the relative density fluctuations are
 \begin{equation}
 \frac{\delta \rho_i}{\rho_i}=\frac {8\pi\, \rho_i}{3\frac{c^5}{\hbar G^2}}=\frac{8\pi}{3} \frac{\rho_i}{\rho_P}    
 \label{delta} 
 \end{equation}
 where $\rho_P=5.1550 \times10^{96} kg/m^3$ is the Planck density.
 Observations of the CMB shows that  $| {\Delta \rho/\rho} |\approx 6 \Delta T/T\approx 6 \times 10^{-5}, $(\cite{white})
 Using equation \ref{delta} we find that the density during the time when the observed $1^\circ$ CMB fluctuations were formed was 
 \begin{equation}
 \rho_{1^{\circ}}\sim0.72\cdot10^{-5}\, \rho_P\approx3.7 \times 10^{91}kg/m^3
 \label{CMB}
 \end{equation}  
 and it will have the size of the eventhorizon
 \begin{equation}
\lambda_{1^{\circ}}= c/H(t_{1^{\circ}})= \frac{c}   {\left( \frac{8 \pi} {3}   G\rho_{1^{\circ}} \right)^{1/2}   }\approx2.1 \cdot 10^{-32} m
\label {lambda}
\end{equation}
 If the same fluctuation is observed at lss with a wavelenght $\lambda_l$ the redshift is 
\begin{equation}
z_{1^{\circ},l}=\frac{\lambda_l}{c} \left( \frac{8 \pi} {3}   G\rho_{1^{\circ}} \right)^{1/2}
\end{equation}
A fluctuation which has the size $1^{\circ}$ at lss  corresponding to $4.2 \times 10^{21} m$ has thus been redshifted by \\
\begin{equation}
z_{1^{\circ},l} \approx \frac{4.2 \times 10^{21} } {2.1 \cdot 10^{-32}}\approx 2. \cdot 10^{53}
\end{equation}
since it was created, corresponding to an efolding of about 122.
Since the density decreases slowly during inflation it is slightly less than $\rho_{1^{\circ}}$ at the end of inflation, so $\rho_{ie} \approx 10^{90}kg/m^3$.
The sum of matter and radiation density when they are equal, at a redshift of 3400 is about $\rho_{rel}\approx 2 \cdot10^{-16} kg/m^3$. The redshift from lss to $z_{eq}$ is $3400/1100$, or a scale efolding of about $1.1$. 
From there on, making the simplifying assumption that dark matter behaves as ordinary matter, the energy density dependence 
on the redshift up to the end of inflation is $\sim z^4$. The redshift from $z_{eq}$ to $z_{ie}$ is  then $z_{eq,ie}=\left(\frac{10^{90}} {2 \cdot 10^{-16}} \right )^{1/4} \approx 2.7\cdot 10^{26}$
corresponding to an efolding of about 61. Thus inflation must from the $1^{\circ}$ fluctuation point contribute an efolding of about $122-1-61= 60$.\\
\section{the power index}
Change in the density during inflation will cause a wavelength dependance of the CMB fluctuations.  
   The density fluctuations can be expressed in terms of a logarithmic power spectrum, 
   \begin{equation}
   \left(\frac{\delta\rho}{\rho}\right)^2=\int P(k) d ln( k)
   \end{equation}
   ,where $k$ is the spatial frequency.
   The power spectrum is usually approximated as
 \begin{equation}
 P(k)\sim k^{n-1}
 \label{ps}
 \end{equation}
 where $n$ is the spectral index. 
 Planck observations \cite{planck} give a spectral index, $n = 0.9649$.
 \begin{figure}
\resizebox{0.85\hsize}{!} {\rotatebox{-90}{\includegraphics {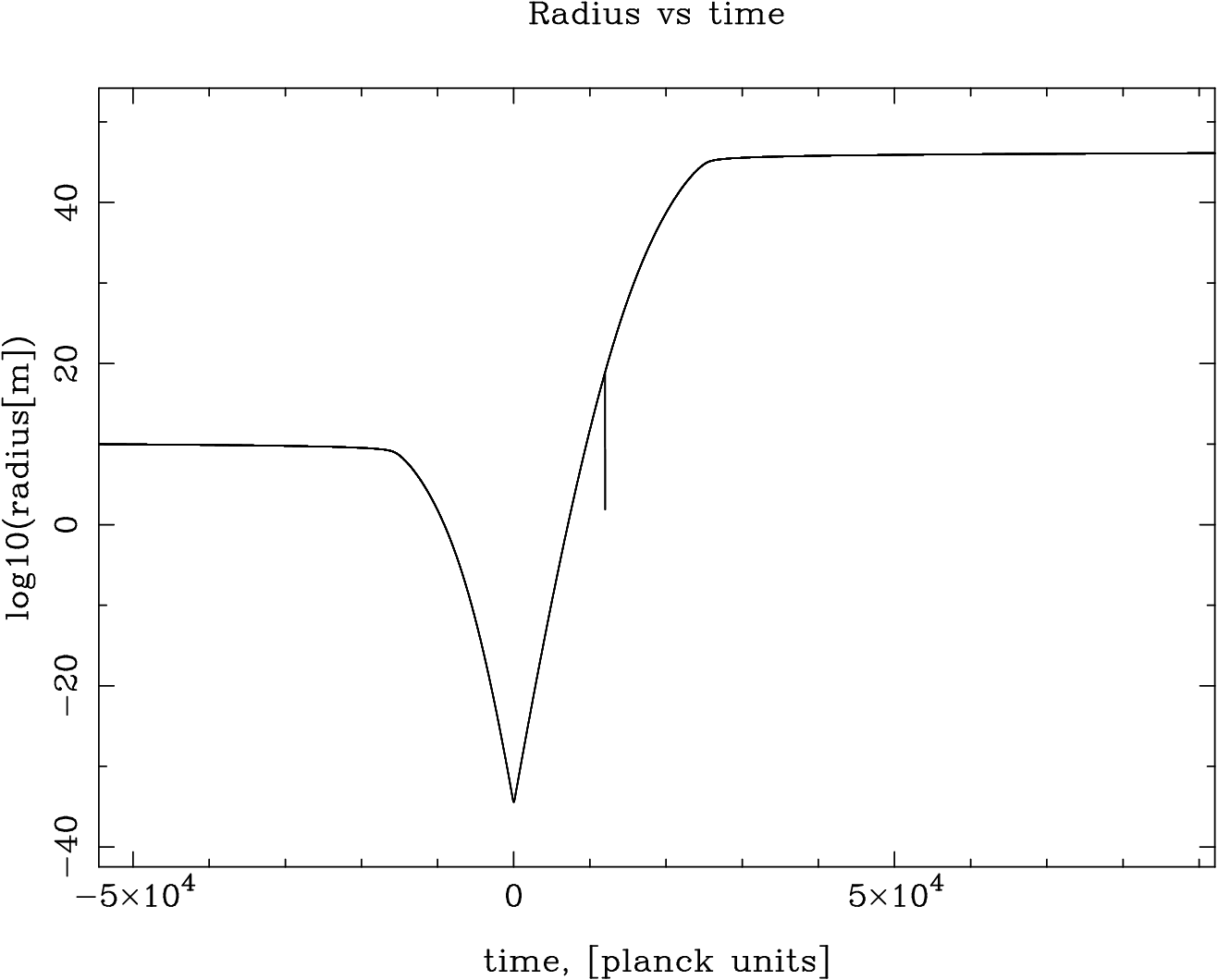}}}
\caption{Radius of the model universe as a function of time. The radius where the $1^\circ$ CMB fluctuations were created is 
noted by the vertical line}
\label{radius}
\end{figure}
\begin{figure}
\resizebox{ 0.85\hsize}{!} {\rotatebox{-90}{\includegraphics {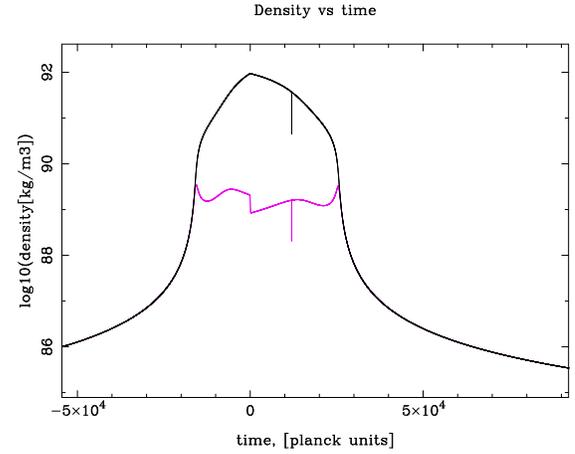}}}
\caption{Total density of the model universe (black) and thermal density (red) as a function of time. The time is zero at the bounce.  
The density where the $1^\circ$ CMB fluctuations were created is noted by the vertical line}
 \label{density} 
\end{figure}
\section{The power index and the density scale index}
One can relate the spectral index n of the power spectrum of density fluctuations to the density scale index m.
\begin{equation}
\rho=\rho_0 \left(\frac{a_0}{a}\right)^m
\end{equation}
\begin{equation}
a \sim \rho^{-1/m}
\end{equation}
The wavelenght of a density perturbation when created is (cf eqn. \ref{lambda})
\begin{equation}
\lambda_i= c/H(t_i)= \frac{c}   {\left( \frac{8 \pi} {3}   G\rho_i \right)^{1/2}   }
\end{equation}
When observed it has the wavelenght 
\begin{equation}
\lambda_{obs}=\frac{a_{obs}}{a_i} \frac{c}   {\left( \frac{8 \pi} {3}   G\rho_i \right)^{1/2}   }
\end{equation}
Thus, ($a_{obs}=const)$
\begin{equation}
 \rho^{1/2}  \sim a^{-1} \lambda^{-1}  \sim k/a
\end{equation}
\begin{equation}
\rho^{1/2-1/m}\sim k
\end{equation}
\begin{equation}
\rho \sim k^{\frac{1}{1/2-1/m}}
\end{equation}
 According to equation \ref{ps} and \ref{delta} ,
 \begin{equation}
 \rho^2 \sim k^{(n-1)}\sim \lambda^{ (1-n)}
 \end{equation}
we then have
\begin{equation}
k^{(n-1)/2} \sim k^{\frac{1}{1/2-1/m}}
\end{equation}
thus
\begin{equation}
n-1 = \frac{2}{1/2-1/m}
\end{equation}
or
\begin{equation}
m=\frac{2(1-n)}{5-n}
\end{equation}
If $n=0.965$ then
$m=0.01735$, i.e. the energy density is nearly scalar and the expansion is nearly exponential.
\section{The density scale index function}
\begin{figure}
\resizebox{0.85\hsize}{!} {\rotatebox{-90}{\includegraphics {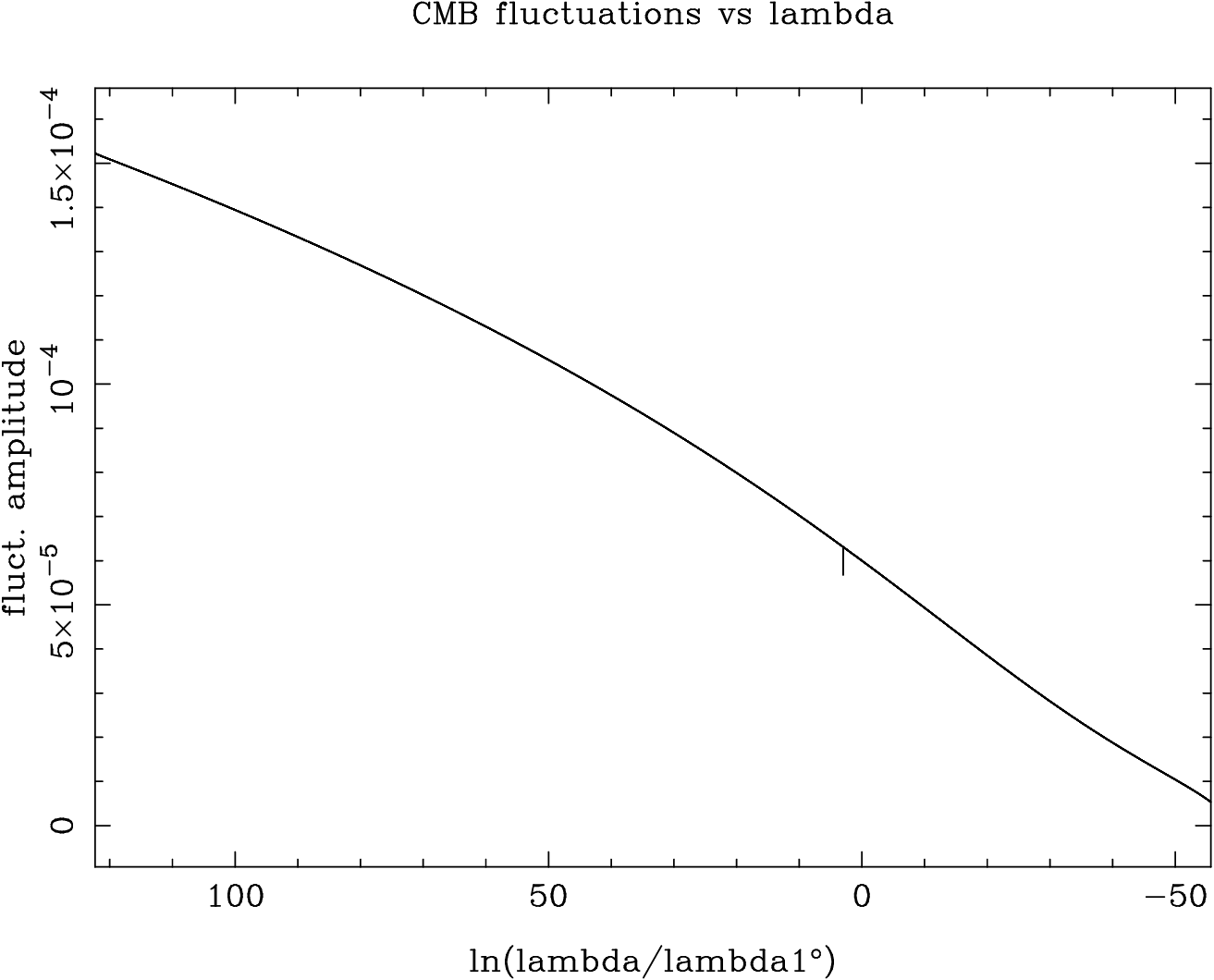}}}
\caption{ Model CMB fluctuations as a function of ln of the wavelenght $lambda/lambda1^\circ=\lambda/\lambda_{1^\circ}$. The vertical line indicates wavelenght corresponding to
 the Hubble radius.}
 \label{D.fluct}
\end{figure}

\begin{figure}
\resizebox{0.85\hsize}{!} {\rotatebox{-90}{\includegraphics {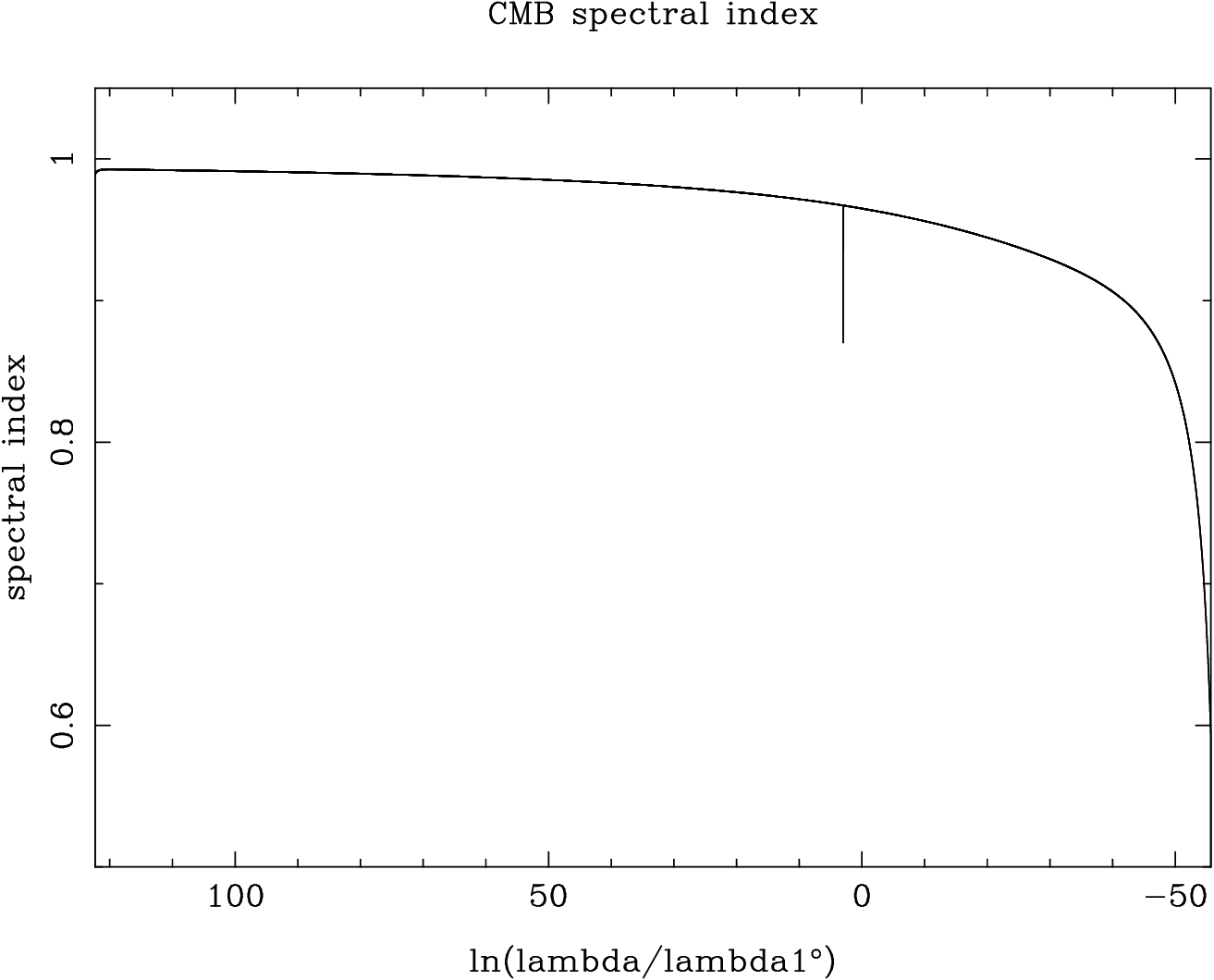}}}
\caption{Spectral index of the model CMB fluctuations as a function of ln of the wavelenght lambda. The vertical line indicates wavelenght corresponding to
 the Hubble radius.}
 \label{index}
\end{figure}

We will now attempt to construct a scale index function  $m(\rho)$ which satisfies the above shown constraints.
\begin{enumerate}
\item $m(\rho)$ must approch zero as $\rho$ approaches the upper density limit
\item $\rho_{1^{\circ}}\approx3.7 \times 10^{91}kg/m^3$\\
\item $m(\rho_{1^{\circ}})\cdot (1- (\rho_{1^\circ}/\rho_m)^{1/2})=0.01735$ \\ assuming $\rho_t(1^\circ) \approx \rho_{1^\circ}$
\item $m(\rho)$ must remain small for densities slightly lower than $\rho_{1^{\circ}}\ $ to make inflation produce 60 efoldings  
\item It must have the limit 4 for low (but relativistic) densities, if indeed, 4 is the correct limit.
\end{enumerate}
When $m(\rho)$ is small the space inflates. Since $m(\rho_{1^{\circ}})$ is small, one may assume that $\rho_{1^{\circ}}$ is fairly close to the upper density limit.
We construct a trial equation
\begin{equation}
m(\rho)= \frac{C_1\cdot  (1\pm (\rho_t/\rho_m)^{1/2}}{1+(\rho/\rho_1)^2}+\frac{C_2\cdot  (1\pm (\rho_t/\rho_m)^{1/2})}{1+(\rho/\rho_{1^{\circ}})^2} 
\end{equation}
$ \rho_{1^\circ}= 3.7 \cdot 10^{91}kg/m^3$ \\
$\rho_1=k_1\cdot\rho_{1^\circ}$\\
$C_2 = k_2\cdot 0.01735 \cdot 2/ (1- (\rho_{1^{\circ}}/\rho_m)^{1/2})$ and \\
$C_1=4-C_2 $\\
$k_1 << 1.0 $, determines the duration of inflation for densities lower than $\rho_{1^{\circ}}$.  $k_2 \approx 1.0$ determines the behaviour 
of the scale index for densities around $\rho_{1^\circ}$.  The limit 4 for the scale index for lower densities might be questioned, see discussion. There are no observational constraints on the  behaviour of the scale index for densities much above $\rho_{1^\circ}$.  It is hard to get densities much higher than $\rho_{1^\circ}$ unless the scale index function returns to a higher value. The suggested function actually doesn't prevent $\rho$ from passing $\rho_m$ but it never occurred in my trials and is not important since the objectiv of this work is to show that the approach is possible and not that it is true in detail. If the density  should approach the Planck density the density fluctuations will get wild (cf eqn.\ref {delta}). We have found that $k_1=0.0180$ and $k_2=0.945$ results in a spectral index of 0.965 at the $1^\circ$ fluctuation and an efolding of 60.0 to the end of inflation and a final inflation density of about $10^{90} kg/m^3$, which is in line with the previous estimates and observational constraints. The evolution of the model universe radius and density is shown in figures \ref{radius} and \ref{density}. Model relative density fluctuations as a function of $ln(\lambda)$   is shown in fig. \ref{D.fluct} and the spectral index of these fluctuations in fig. \ref{index}.
 \section{limits on the spatial curvature}
 The density parameter is defined as
\begin{eqnarray}
\Omega \equiv \rho/ \rho_c=\frac{8\pi G\rho}{3 H^2}  \\
\rho_c=\rho_{critical} 
\end{eqnarray}
and the curvature density parameter 
\begin{equation}
\Omega_k =1-\Omega 
\end{equation}  
We have then for todays radius of the universe
\begin{equation}
R_U = \frac{c}{H \,(-\Omega_k)^{1/2}}
\end{equation}
Observations (\cite {okouma}) limit $|{\Omega _k}|$  to $\leq 10^{-3}$.
The hubble radius $R_H=c/H\sim1.4\cdot 10^{10}$ ly.
The lower limit of the radius of the present universe is
$R_U \ge 32 \times R_H=4.6\cdot 10^{11}$  ly. Our bounce model universe gives a comoving radius of about $10^{53}$ ly, which is relevant only to the extent
 that it is larger than the above limit. \\
\section{discussion}
My first thoughts on the possibility of an upper density limit were strengthened when I read the important but largely forgotten papers by Gliner \cite{gliner} and Sacharov \cite{sakharov}, 
which resulted in the publication\cite{rydbeck}. In time I came to understand that there is a fundamental problem with the standard inflation scenario in that it is 
lacking a preferred frame of reference required for particle production (thanks L. Susskind for advice on this matter). I then realized that it should be possible to modify my bounce model so that it had a density 
dependant equilibrium relation between relativistic matter-radiation and vacuum energy density, thereby maintaining a preferred restframe and still support the bounce and inflation. If such a relation continued as densities became lower and eventually reached todays densities, it might resolve the "Hubble tension" 
\cite{marc} and also explain the "coincidence problem" \cite{velten}. It could perhaps as well put a new light on the nature of vacuum energy\cite{luca},\cite{camero}.  One might in this vein note that the universe is again inflating although at a much lower rate.\\
While we have assumed a smooth density distribution in e.g. a stellar center, the first phase involving a collapse process may become very fractured. The vacuum energy density  relation to particle-radiation energy density will involve dynamics a subject on which one can find a number of publications, eg \cite{dutta},\cite{sola},\cite{anisimov}. Thus, if our model for a bounce  inflation is
anywhere near a real inflation event, it is of course still very simplified.

\end{document}